\documentclass{article}
\usepackage[applemac]{inputenc}
\usepackage[english]{babel}
\usepackage{geometry}
\usepackage{amssymb,amsmath}
\usepackage{graphicx,color}
\usepackage{mathrsfs}
\usepackage{hyperref}
\usepackage{cite}

\newcommand{\kB}{\ensuremath{k_{\mathrm{B}}}} 
\newcommand{\ave}[1]{\ensuremath{\left \langle {#1} \right \rangle}} 
\newcommand{\eps}{\ensuremath{\varepsilon}} 

\newcommand{\om}{\ensuremath{\omega}}

\newcommand{  \p   }   { \ensuremath{      \big{(}          }               }
\newcommand{  \q   }   {  \ensuremath{      \big{)}        }                }
\newcommand{\g}[1]{``{#1}''}

\begin{document}
\begin{titlepage} 
\title{On the ideal gas law}

\author{%
Jacques \textsc{Arnaud}
\thanks{Mas Liron, F30440 Saint Martial, France},
Laurent \textsc{Chusseau}
\thanks{Institut d'Électronique du Sud, UMR n°5214 au CNRS, Université Montpellier II, F34095 Montpellier, France},
Fabrice \textsc{Philippe}
\thanks{LIRMM, UMR n°5506 au CNRS, 161 rue Ada, F34392 Montpellier, France}
}
\maketitle

\begin{abstract} 
The air density on earth decays as a function of altitude $z$ approximately according to an $\exp(-w\,z/\theta)$-law, where $w$ denotes the weight of a nitrogen molecule and $\theta=\kB T$ where $k_B$ is a constant and $T$ the thermodynamic temperature. To derive this law one usually invokes the Boltzmann factor, itself derived from statistical considerations. We show that this (barometric) law may be derived solely from the democritian concept of corpuscles moving in vacuum. We employ a principle of simplicity, namely that this law is \emph{independent} of the law of corpuscle motion. This view-point puts aside restrictive assumptions that are source of confusion. Similar observations apply to the ideal-gas law. In the absence of gravity, when a cylinder terminated by a piston, containing a single corpuscle and with height $h$ has  temperature $\theta$, the average force that the corpuscle exerts on the piston is: $\ave{F}=\theta/h$. This law is valid at any temperature, except at very low temperatures when quantum effects are significant and at very high temperatures because the corpuscle may then split into smaller parts. It is usually derived under the assumption that the temperature is proportional to the corpuscle kinetic energy, or else, from a form of the quantum theory. In contradistinction, we show that it follows solely from the postulate this it is independent of the law of corpuscle motion. On the physical side we employ only the concept of potential energy.  A consistent picture is offered leading to the barometric law when $w\,h\gg\theta$, and to the usual ideal-gas law when $w\,h\ll\theta$. The mathematics is elementary. The present paper should accordingly facilitate the understanding of the physical meaning of the barometric and ideal-gas laws.

\end{abstract}

\end{titlepage}

\newpage

\section{Introduction}\label{introduction}

The purpose of this paper is to show that the barometric law and the ideal-gas law may be obtained on the sole basis of the democritian model according to which nature consists of corpuscles moving in a vacuum, plus a principle of simplicity: namely that these fundamental laws are \emph{independent} of the law of corpuscle motion (non-relativistic, relativistic, or otherwise: see appendix \ref{motion}). The temperature $\theta$ enters into the ideal-gas law and the expression of the gas internal energy solely for dimensional reasons. We show from the general expressions of the gas internal energy and of the force (or pressure) that the heat delivered by the gas is $\theta \,dS$, an expression of the entropy $S$ being given. This result enables us to prove that the formally-introduced temperature $\theta$ coincides, to within some arbitrary constant factor, with the thermodynamic temperature $T$. Indeed, we recover for ideal gases the general Carnot result asserting that the maximum efficiency of thermal engines is: $1-\theta_l/\theta_h$, where $\theta_l$ denotes the cold bath temperature and $\theta_h$ the hot bath temperature. The reader may feel that our statement that the above invariance principle \emph{implies} the barometric and ideal-gas laws, without anything else, is quite surprising. Yet, we hope that we can convince him/her that this is indeed the case. 

Let us emphasize that our goal is to derive the barometric and ideal-gas laws from first principles, only conservation of potential energy being assumed. We do not use the concept of kinetic energy, nor do we postulate any law of corpuscle motion. Accordingly, a given potential $\phi(z)=w\,z$ for a weight $w$ does not imply any specific law of motion $z(t)$.

\paragraph{Concepts relating to heat since the antiquity:}

Democritus, who lived about 300 years B.C., described nature as a collection of corpuscles that cannot be split, moving in vacuum. These corpuscles differ from one-another in form, position and weight. In the case of a gas, interaction between corpuscles may often be neglected, but they collide with the container's walls. Platon \cite{hecht} ascribed heat to corpuscular motion: \g{Heat and fire are generated by impact and friction, but that's motion}. Much later, Francis Bacon (1561-1626) wrote: \g{The very nature of heat is motion, and nothing else}. This view-point is more explicit in Daniel Bernoulli writing (1738): \g{Gas atoms are moving randomly, and pressure is nothing else but the impact of the atoms on their container walls}. Lastly, Carnot introduced energy considerations \emph{circa} 1830: \g{Heat is nothing but motive power, or rather another form of motion. When motive power is destroyed, heat is generated precisely in proportion of the motive power destroyed. Likewise, when heat is destroyed, motive power is generated }. We will employ the law of conservation of potential energy, well known since the antiquity from cords and pulleys experiments.

\paragraph{Experimental results relating to air:}

The first accurate experiments relating to gases are tied up to the invention of the thermometer by Galileo and the barometer by his assistant Torricelli. Then, to Pascal experiments on atmospheric pressure. Pascal ascribed the diminution of the height of a mercury column as a function of altitude to the reduction of the weight of the air above the barometer. It was later shown that the pressure decays exponentially.

Let us recall the crucial experiments performed in the seventeenth century concerning the properties of air. Air, consisting mostly of di-atomic nitrogen, may be viewed as an ideal gas. When a tight box contains some amount of air, the volume-pressure product is a constant at room temperature, a law enunciated for the first time by Boyle in 1660: \g{Pressure and expansion are in reciprocal proportions}. Boyle employed a J-shaped glass tube, with the sealed small side full of air, and the other full of mercury. The left-side height was a measure of volume and the right-side height a measure of pressure. Subsequent experiments have shown that this law is applicable at any constant temperature, for example at various liquid boiling temperatures, within some experimental range. We call \g{generalized Boyle law} the expression: $\textsf{V}(\textsf{P},\theta)=f(\theta)/\textsf{P}$, where $\textsf{P}$ represents pressure, $\textsf{V}$ volume, and $f(\theta)$ some temperature measure.

From an experimental standpoint we could define temperature as the pressure relating to some given amount of matter contained within some fixed volume. As the temperature gets higher the pressure increases. This pressure may be used to define $\theta$. Of course, different temperature scales would be obtained for different substances, but such thermometers may be calibrated one against another because temperatures tend to equalize in equilibrium. Rarefied helium may be described with great accuracy as a collection of independent corpuscles, except perhaps at very low temperatures when quantum effects become significant and at very high temperatures when the helium atoms may get ionized. The theory presented in this paper shows on the basis of the corpuscular model that $\theta$, initially introduced formally from dimensional considerations, coincides with the thermodynamic temperature. This is the temperature that enters in the expression of thermal-engine efficiencies.

Gay-Lussac has shown in 1802 that, at atmospheric pressure, the volume increment of various gases from freezing to boiling water temperatures is 37.5\%\cite{holbrow}. Appropriate gas thermometers enabled experimentalists to establish the proportionality of volume and temperature at constant pressure. This measurement was subsequently made at various pressures, for exemple at various altitudes. The generalized Gay-Lussac law may be enunciated as follows: The two-variable function $\textsf{V}(\textsf{P},\theta)=\theta \,g(\textsf{P})$, where $g(\textsf{P})$ is some unknown function of pressure. Comparison of the generalized Boyle and Gay-Lussac laws shows that: $\textsf{P}\,\textsf{V}=\theta\, h(N)$, where $h(N)$ defines the amount of gas considered.

Let us emphasize that the empirical Gay-Lussac law makes sense only if one specifies which thermometer is  employed. One may employ a gas thermometer from a selected gas such as helium in two ways. One method consists of defining the temperature as the cylinder height (or volume) at a fixed pressure, for example at the standard atmospheric pressure. In the following, we assume that a second method is being employed instead: the temperature is defined as the force that must be exerted on the piston to maintain the height at a fixed value, for example one meter, as said above. If the Gay-Lussac experiment were applied to a gas identical to the gas employed in the thermometer (helium in our example) the fact that pressure is proportional to temperature would be obvious. The importance of the Gay-Lussac experiment is that the proportionality law is found to be valid for any gas. 

However, it was subsequently discovered that the Gay-Lussac proportionality law is reasonably accurate only at very small pressures. The theoretical reason that explains this observation is that, at low pressures, the gas molecules of the tested gas and those of the thermometer gas may both be considered as independent non-interacting corpuscles (see below).  

In 1803, Dalton, on the basis of his studies of chemical compounds and gaseous mixtures suggested that matter consists of atoms of different masses that combine in simple ratios. He discovered the partial-pressure law according to which the total pressure exerted by a gas mixture is equal to the sum of the pressure that each one of the gases would exert if it occupied the full volume alone. Finally, in 1811, Avogadro concluded that equal volumes of gases at the same temperature and pressure contain the same number of molecules (or corpuscles). This entails that $\textsf{P}\,\textsf{V}/\theta$ is proportional to $N$, now interpreted as the number of corpuscles. One calls \g{Avogadro number}, $N_A$, the number of corpuscles contained in 0.0224 cubic meters of gas in standard conditions. In 1865, Loschmidt established from a measurement of the air viscosity that $N_A$ is on the order of $10^{23}$. Many other methods have been employed since then for that purpose, such as Brownian motion.

On empirical grounds, the ideal-gas law may therefore be written as:
\begin{align}\label{id}
\textsf{P}\,\textsf{V}=N\,\theta
\end{align}
where $\textsf{V}$ denotes the volume, $\textsf{P}$ the pressure, $N$ the number of corpuscules, and $\theta\equiv \kB T$ the temperature. 
 
The ideal-gas law has been partly explained on the basis of a kinetic theory by Waterston \cite{waterston} in 1843, the kinetic theory being based itself on non-relativistic mechanics. The next important theoretical discovery is due to Boltzmann, see Section \ref{usual}. In subsequent sections we recall the basic assumptions on which rest the usual proofs of the barometric and ideal-gas laws. Then we present our model. 

\section{Usual kinetic and statistical theories}\label{usual}

A recent reference\cite{kinetic} lists the assumptions on which the gas kinetic theory rests. Some of them express the democritian hypothesis and are indeed essential. The usefulness of the others, listed below, however, may be questioned:
\begin{enumerate}

\item Gases consist in corpuscles having non-zero mass.

\item The corpuscles are quickly moving.

\item They are perfectly spherical and elastic.

\item The average kinetic energy depends only on the system temperature.

\item Relativistic effects are negligible.

\item Motion laws are time reversible. 

\item The number of molecules is so large that a statistical treatment is appropriate.\label{sept}
\end{enumerate}
\noindent Comment: As we shall show, none of the above assumptions are needed. It suffices that the (perhaps unique) corpuscle be in thermal contact with the ground.\\

We sketch in the present section the most usual derivations of the barometric and ideal-gas laws to remind the readers of the underlying assumptions. Note that the barometric law may be obtained from the ideal-gas law, and conversely, if one postulates that weightless plates may be introduced or removed at will in the gas at various altitudes. But this postulate is at best plausible. In an interesting paper, Norton \cite{Norton:2006} derived the ideal-gas law from the barometric equation. However, the latter involves the Boltzmann factor, which requires other physical considerations (see below), while in the present paper this factor comes in naturally, that is, for purely mathematical reasons.

The barometric law is usually viewed as a straightforward consequence of the Boltzmann factor: the probability that a corpuscle has energy $E$ is proportional to: $\exp(-E/\theta)$ where $\theta=\kB T$ and the energy $E=w\,z$, where $w$ denotes the corpuscle weight (e.g., the weight of a di-atomic nitrogen molecule) and $z$ the altitude. Hence the exponential decay. 

The derivation of the Boltzmann factor itself is based on a quantization of the energy\footnote{This concept was introduced by Boltzmann who, however, considered only the limit in which the difference between successive energies is arbitrarily small.}, and the postulated equi-probability of the micro-states. Let the discrete (non-degenerate) energy levels be denoted by $\eps_1,\,\eps_2,...$. If distinguishable corpuscles are distributed among the energy levels, with $n_1$ corpuscles in level 1, $n_2$ corpuscles in level 2, and so on, the number of ways of doing that is inversely proportional to: $n_1!\,n_2!...$. It is postulated that this number reaches its maximum value at equilibrium under the constraint that $n_1+n_2+...=N$, the total number of corpuscles, and $n_1\eps_1+n_2\eps_2+...=E$, the total energy. In the limit of large $N$ values, one finds that: $n_i\propto\exp(-\eps_i/\theta)$, for some $\theta$-value that depends on $N$ and $E$. Even though physicists are now-a-day very familiar with that procedure, it is not so easy to explain it to students. Besides, it rests on a number of assumptions. 

The traditional derivation of the ideal-gas law, on the other hand, is based on non-relativistic mechanics \cite{Feynman:1963}.  For a one-dimensional model, one considers a corpuscle moving back and forth between two plates separated by a distance $h$, one of them playing the role of a piston. If $v$ denotes the speed of a corpuscle, an impact on a plate delivers to it an impulse $2\,m\,v$ where $m$ denotes the corpuscle mass, and this occurs every $2h/v$ time units. It is concluded that the force $F$ exerted on the piston is: $2\,m\,v/(2h/v)=m\,v^2/h$, that is: $F\,h=2\,K$, where $K=\frac{1}{2}m\,v^2$ denotes the kinetic energy. It is recognised that there may be a distribution of kinetic energies \cite{Feynman:1963}. Postulating that the temperature $\theta$ is proportional to the average kinetic energy one obtains for the average force the ideal-gas law: $\ave{F}h\propto \theta$. Alternatively, one may quantize the corpuscle wave-function and employ the Bolzmann factor \cite{arnaud2}.

The procedure described above has been generalized to relativistic motion ($\kB T\sim mc^2$). The same ideal-gas law is valid at any temperature (within the corpuscular model). Our thesis is that the ideal-gas law has simply nothing to do with the law of corpuscle motion, and that it is therefore not surprising that it holds for both the Galileo and Einstein laws of motion. The assumption that temperature is proportional to the average kinetic energy cannot possibly be derived from first principles since it is only an approximation acceptable when $\kB T\ll mc^2$. These are some of the reasons why we feel that the traditional proofs are unsatisfactory. An alternative is offered below.

\section{The barometric law}\label{barom}

We are considering an unit-area cylinder with vertical axis in uniform gravity, resting on the ground ($z=0$) at some temperature. We consider only motion along the vertical $z$-axis, denoted in general by $z=z(t;E)$, where $t$ denotes time. The corpuscle energy is defined as: $E\equiv w\,z_m$ where $w$ is the corpuscle weight and $z_m$ the maximum altitude. In the following, some regularity of the $z(t)$-function is assumed, but no specific law is presumed, except in examples. We set for convenience $t=0$ at the top of the trajectory, that is: $z(0)=z_m$, $z'(0)=0$, where a prime denotes a derivative with respect to $t$.

\begin{figure}
\centering
\includegraphics[width=0.6\columnwidth]{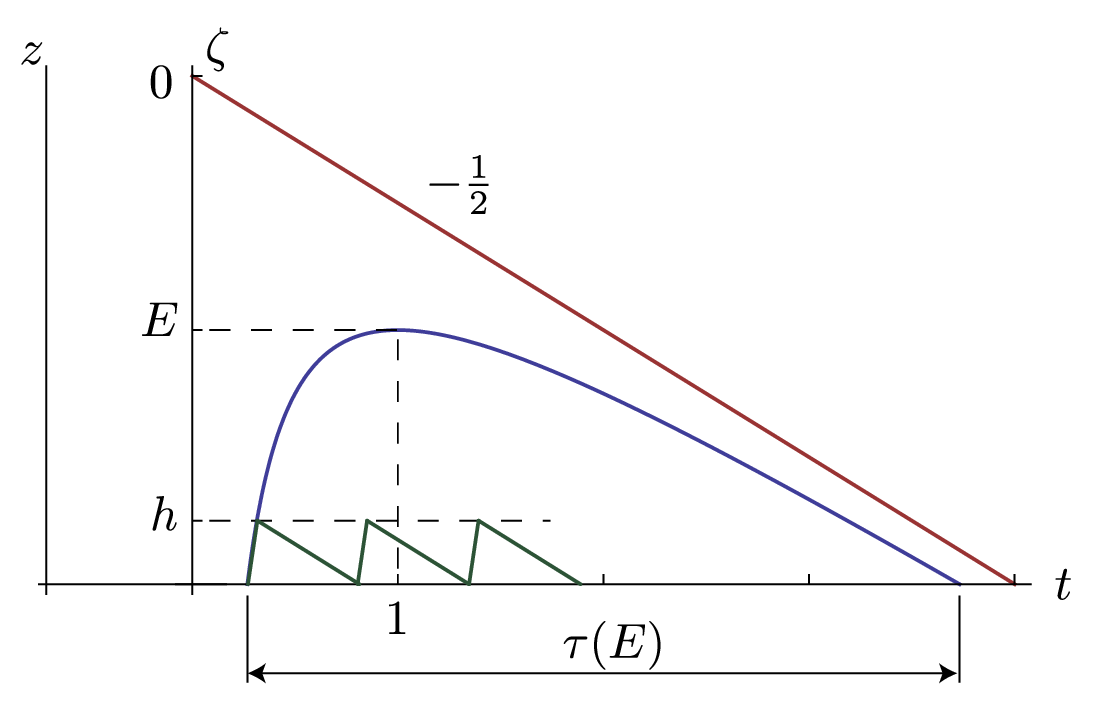}
\caption{The curve represents the motion $z=z(t)$ of a corpuscle of weight unity submitted to static uniform gravity. Note that the results presented in this paper do not depend on the law of motion, which needs not be an even function of time. (The particular curve shown refers to a relativistic law of motion based on a clock synchronisation different from the usual (Einstein) one. Namely, we suppose that clocks at various altitudes $z$ are synchronized by a light pulse emitted from $z=0$ at $t=0$. Then the apparent speed of light is infinite upward and equal to $c/2$ downward. The asymptote is $\zeta\equiv z-z_m=-1/2t$. We give these details because the reader may be puzzled by the curve represented on this figure. However, almost \emph{any} other curve would serve our purposes as well).  The maximum height $z_m$ reached above the ground level $z=0$ is the corpuscle energy $E$. $\tau(z_m)$ represents the motion period. A reflecting plane (piston) is shown at the height $z=h$ as a dotted line, the corpuscle bouncing alternately off the two planes. From a simple inspection of the figure on sees that the period becomes in that case: $\tau(z_m)-\tau(z_m-h)$.}
\label{fig}
\end{figure}

Consider a single period of corpuscle motion as shown in Fig. \ref{fig}. Let the corpuscle distance from the top of its trajectory  be denoted $Z\equiv z_m-z≥0$, at times $t_1$ and $t_2≥t_1$. We call\footnote{We do not assume that $z(t)=z(-t)$ or $t_2+t_1=0$. Asymmetric laws of motion occur if one employs clock synchronisation rules different from the one proposed by Einstein. For example, if a light pulse emitted from $z=0$ at $t=0$ propagating upward is employed to synchronise clocks located at different altitudes, the apparent upward speed of light is, by this convention, infinite. The downward speed of light is then $c/2$ if $c$ denotes the Einstein speed of light, so that the two-way speed of light remains equal to $c$, in agreement with very precise experiments. The present anisotropy refers to a change of formalism, not of physics. It is of some importance that the laws discussed in this paper do not depend on such conventional changes.}: \g{time interval} $\tau(Z)\equiv t_2-t_1$. Because gravity is static and uniform (that is, independent of altitude and time) this $\tau$-function depends only on $Z$. As an example, for non-relativistic motion: $Z(t)=\frac{1}{2}g\,t^2$, with $w\equiv m\, g$ where $m$ denotes the corpuscle mass and $g$ the gravitational acceleration. In that example: $\tau(Z)=2\sqrt{2\,Z/g}\propto \sqrt{Z}$, $g$ being a constant.

The period of motion of a corpuscle bouncing off the ground ($z=0$) without any loss of energy (rigid walls and negligible gas friction), and having energy $E\equiv w\,z_m$, is according to the above definitions: $\tau(z_m)$. On the other hand, the time spent by the corpuscle above some $z$-level is obviously zero if $z>z_m$, and equal to: $\tau(Z)\equiv\tau(z_m-z)$ if $z≤z_m$. In the latter case, the fraction of time during which the corpuscle is above $z$ is therefore: $\tau(z_m-z)/\tau(z_m)$, as suggested on the figure \ref{fig}. 

We now suppose that the ground on which the corpuscle is bouncing off has been heated to some temperature (the concept of temperature will be precisely defined later on). This means that the ground level ($z$=0) is not perfectly static as assumed above, but instead is quivering. Concretely, the groung level could be moving up and down according to some zero-mean fluctuation of small amplitude so that, upon impacting on the ground, the corpuscle may loose energy (when the ground level is moving downward), or gain energy (when the ground level is moving upward). We will not need the explicit form of this fluctuation. We only assume that the amplitude of that fluctuation is so small that the corpuscle energy does not vary significantly over many periods. Yet it may evolve slowly. The energy distribution $\omega(E)$ refers to averages over arbitrarily long times, and is presently unknown; it will be determined by enforcing the condition that the law of interest (presently the barometric law) does not depend on the corpuscle law of motion.

The fractional time during which the corpuscle is above some $z$-level is, according to the previous considerations and remembering that $z_m=E/w$ where $w$ is a constant: 
\begin{align}\label{barx}
\mathcal{A}(z)=\frac{\int_{wz}^\infty \,dE\,\om(E)\,\tau(E/w-z)/\tau(E/w)}{\int_0^\infty dE\,\om(E)}.
\end{align}
The lower limit of the integral in the numerator is $wz$ since the fractional time is equal to zero when $E≤wz$.

We now argue that $\om(E)$ \emph{must be}: $\exp(-E/\theta)\tau(E/w)$, where $\theta$ is a constant having the dimension of energy as is required by the fact that the argument of an exponential is dimensionless. First, let us introduce this distribution law in \eqref{barx}. We obtain:
\begin{align}\label{barr}
\mathcal{A}(z)&=\frac{\int_{wz}^\infty \,dE\,\exp(-E/\theta)\tau(E/w-z)}{\int_0^\infty dE\,\exp(-E/\theta)\tau(E/w)}\nonumber\\
&=\frac{\exp(-w\,z/\theta)\int_{wz}^\infty \,dE\,\exp(-(E-wz)/\theta)\tau(E/w-z)}{\int_0^\infty dE\,\exp(-E/\theta)\tau(E/w)}\nonumber\\
&=\frac{\exp(-w\,z/\theta)\int_{0}^\infty \,dE'\,\exp(-E'/\theta)\tau(E'/w)}{\int_0^\infty dE\,\exp(-E/\theta)\tau(E/w)}\nonumber\\
&=\exp(-w\,z/\theta).
\end{align}
On the third line, we have used as an integration variable $E'\equiv E-wz$ instead of $E$. The last line follows from the fact that $E,E'$ are dummy variables, so that we may replace $E'$ by $E$. Note that, even though we have introduced integral signs, no integral has been performed.

The distribution $\om(E)$ introduced above is the one that leads to a result (last line of \eqref{barr}) independent of the particular form of the $\tau(Z)$-function, and therefore of the law of motion. For a purely mathematical reason, the term: $\tau(E/w)$ must be there to cancel a similar term in the denominator of the expression of the fractional time. On the other hand, the only function of $u\equiv E/\theta$ that may cause the integral in the numerator to go from 0 to infinity and cancel out with the integral in the denominator is: $\exp(-u)$, the argument being defined only to within a constant factor. In order to obtain the energy distribution $\om(E)$, one would need to know the function $\tau(Z)$. But, remarkably, the energy distribution is not needed explicitly.

From a physical standpoint, the energy distribution may be written as $\exp(-E(f)/\theta)$, where $f$ denotes the action, equal to the $z(t)$ motion area for one period, and $df/dE=\tau(E/w)$. In quantum mechanics the action $f$ is set equal to an integer (1,2..., ignoring a small constant) times the reduced Planck constant $\hbar$. The term $\exp(-E/\theta)$, usually referred to as the \g{Boltzmann factor}, enters here solely by mathematical necessity\cite{arnaud2}.

The gas density, defined as the probability that the corpuscle be located between $z$ and $z+dz$, divided by $dz$, is: $\rho(z)=-d\mathcal{A}(z)/dz=(w/\theta)\exp(-wz/\theta)$. This is the barometric law. Since $w$ and $\theta$ are constant, the density decays exponentially as a function of altitude $z$. In the earth atmosphere the density of di-atomic oxygen decays faster than the density of di-atomic nitrogen because the weight of an oxygen atom exceeds that of a nitrogen atom in the ratio $\approx$ 16/14. 

We have supposed above that the weight $w$ is independent of altitude. A more general formulation is given in appendix.

\section{Potential energy}\label{pot}

We will only need the gas \emph{internal energy} $U$. However, it is of some pedagogical interest to separate out from $U$ a term that may be called the potential energy. This term is directly calculated in the present section. The potential energy of a corpuscle of weight $w$ located at altitude $z$ is defined as $w\,z$. The gas potential energy, according to the above expression of the density, is therefore: $\mathcal{P}=w\int_0^\infty dz\,z\,\rho(z)=w\int_0^\infty dz\,\mathcal{A}(z)=\theta$, using \eqref{barr}. 

When the corpuscle motion is restricted to: $0<z≤h$ one can show that the density $\rho(z)$ is unaffected in that range of $z$, as is discussed in an appendix. It must be normalized, though, so that the integral of $\rho(z)$ from 0 to $h$ be unity. A simple integration of $w\,z\,\rho(z)$ from $z=0$ to $z=h$ then gives the potential energy for any $h$-value:
\begin{align}\label{po}
\mathcal{P}= \theta-\frac{w\,h}{\exp(w\,h/\theta)-1} .
\end{align}
According to that expression $\mathcal{P}$ tends to $\theta$ monotonically if $w\,h\to\infty$. It tends to 0 when $w\,h$ tends to 0, that is in the absence of gravity. This is what is usually assumed when one refers to the ideal-gas law. In that case, as shown below, the internal energy depends only on temperature. 

\section{Internal energy}

The gas internal energy $U$ is the average value of $E$, if only motion along the $z$-axis is considered (note that the gravitational energy is accounted for in $U$). It can be evaluated by the above method and is found to be given by the sum of the potential energy in \eqref{po} and a term that depends on $\theta$ but not on $h$. 

The explicit expression of $U\equiv \ave{E}$ is, omitting details that are similar to those in section \ref{barom}:
\begin{align}\label{uuu}
U&=\frac{\int_0^{w\,h} dE\, E\exp(-E/\theta)\tau(E/w)+\int_{w\,h}^\infty dE\, E\exp(-E/\theta)\p\tau(E/w)-\tau(E/w-h)\q}{\int_0^{w\,h} dE\, \exp(-E/\theta)\tau(E/w)+\int_{w\,h}^\infty dE\, \exp(-E/\theta)\p\tau(E/w)-\tau(E/w-h)\q}\nonumber\\
&=\mathcal{P}+\mathcal{K},
\end{align}
where the potential energy $\mathcal{P}$ is given in \eqref{po} and:
\begin{align}\label{ken}
\mathcal{K}=\frac{\int_0^\infty dE\, E\exp(-E/\theta)\tau(E/w)}{\int_0^\infty dE\, \exp(-E/\theta)\tau(E/w)}-\theta 
\end{align}
depends only on temperature.
 
The explicit form of $\mathcal{K}$ requires the knowledge of the $\tau(Z)$-function and therefore of the corpuscle law of motion. In the special case of non-relativistic motion, for example, we have $\tau(Z)\propto \sqrt{Z}$, see Section \ref{barom}. At that point, Integration is needed. It gives: $\mathcal{K}=\theta/2$, or more precisely: $\mathcal{K}=\kB T/2$, a well-known result. In Physics text-books $\mathcal{K}$ is called the average kinetic energy. This interpretation, however, is not needed here: $\mathcal{K}$ is simply the term that remains when the potential energy is subtracted from the internal energy.

\section{Force exerted on the piston by a corpuscle of energy $E$}

We now treat the ideal-gas law by similar methods. We are considering again a unit-area cylinder with vertical axis in uniform gravity, resting on the ground ($z=0$) at some temperature. A tight piston can move in the vertical $z$ direction. The cylinder height is denoted by $h$ and contains a single corpuscle of weight $w$. In our one-dimensional model, the pressure $\textsf{P}$ corresponds to the average force $\ave{F}$, the volume $\textsf{V}$ to the height $h$, and $N=1$. Our result provides the ideal-gas law in a generalized form, taking into account gravity. In that case, the pressure varies as a function of altitude. More precisely, the force exerted by the corpuscle on the lower end of the cylinder exceeds the force exerted on the upper end (or piston) by the corpuscle weight. But in the absence of gravity, the forces exerted on both ends are the same. 

We are introducing (static and uniform) gravity, not so much for the sake of generality, but because this helps clarify the concept of corpuscle energy: the corpuscle energy is defined as the maximum altitude that the corpuscle would reach in the absence of the piston, multiplied by the corpuscle weight. Precisely, the maximum height reached by a corpuscle of weight $w$ and energy $E$ in the absence of a piston is: $z_m\equiv E/w$, the corpuscle bouncing elastically from the ground, that is, without any loss or gain of energy. 

Consider first the case where $h$ is infinite, that is, in the absence of a piston. The time period is denoted as before $\tau(Z)$ with $Z=z_m$. The average force exerted on the ground, equal to the corpuscle weight $w$, is the product of the impulse $i$ and the number of impulses per unit time. Thus $w=i/\tau(z_m)\longrightarrow~~i=w\,\tau(z_m)$. In other words, the impulse transmitted to a plane when the corpuscle impacts on it is the product of the corpuscle weight and the motion period.

If the plate is located at $z=h$, we have $Z=z-z_m$ and the impulse is: $i_h=w\,\tau(z_m-h)$. When the corpuscle is moving back and forth beween the planes at $z=0$ and $z=h$ (instead of being located above $h$) the impulse is just opposite to $i_h$. It is therefore in absolute value equal to $w\,\tau(z_m-h)$.

Next, we introduce a rigid plane at $z=h$, viewed as a piston, and consider a corpuscle bouncing on the $z=0$ and $z=h$ planes alternately. The time period becomes: $\tau(z_m)-\tau(z_m\,-h)$, as one can see from inspection of the figure. We call $F$ the force exerted on the $z=h$ rigid plane, averaged over a time period. It follows from the previous expressions that:
\begin{align}\label{forcebis}
F&=0\qquad &z_m≤h\nonumber\\
F&=\frac{i_h}{\tau(z_m)-\tau(z_m-h)}=w\frac{\tau(z_m-h)}{\tau(z_m)-\tau(z_m-h)} \qquad &z_m>h.
\end{align}

\section{Average force}\label{average}

As a consequence of the slight quivering of the cylinder lower end (thermal motion), the corpuscle energy $E$ slowly varies in the course of time. The force $F$ just defined must be weighed by some energy distribution $\om(E)$ in such a way that the average force $\ave{F}$ be \emph{independent} of the corpuscle equation of motion, and thus of the $\tau(.)$-function. In the limit where $w\,h\ll \theta$, a condition that amounts to ignoring gravity, we obtain the ideal-gas law in the form: $\ave{F}=\theta/h$, where $\theta$ is a quantity having the dimension of an energy. We later on prove that $\theta$ is the thermodynamic temperature.

The above condition obtains from \eqref{forcebis} if and only if one selects the following energy distribution:
\begin{align}\label{pond}
\om(E)=&\exp(-E/\theta)\tau(E/w)\qquad &E≤w\,h\nonumber\\
\om(E)=&\exp(-E/\theta)\p\tau(E/w)-\tau(E/w-h)\q \qquad &E>w\,h,
\end{align}
where $\theta$ has the dimension of an energy. Indeed, the average force becomes, using \eqref{forcebis} and \eqref{pond}:
\begin{align}\label{moy}
\ave{F}&=w\frac{\int_{w\,h}^\infty dE\, \exp(-E/\theta)\tau(E/w-h)}{\int_0^{w\,h}\, dE\, \exp(-E/\theta)\tau(E/w)+\int_{w\,h}^\infty dE\, \exp(-E/\theta)\p\tau(E/w)-\tau(E/w-h)\q}\nonumber\\
&=\frac{w}{\exp(w\,h/\theta)-1}.
\end{align}
In the above integrals going from $w\,h$ to $\infty$ we have replaced $\exp(-E/\theta)$ by $\exp(-w\,h/\theta)\exp(-(E-w\,h)/\theta)$ and introduced the variable $E'\equiv E-w\,h$, so that all the integrals go from zero to infinity and cancel out. Note that, as in Section \ref{barom}, no integration has been actually performed. 

The corpuscles being independent, for an arbitrary collection of $N$ corpuscles having the same weight the force is multiplied by $N$. In the case of zero gravity $w$=0 or more precisely: $w\,h\ll \theta$. The above expression then gives: $\ave{F}h=\theta$. Thus we have obtained the ideal-gas law: $\ave{F}h=N\,\theta$. 

The expressions given earlier for the average force $\ave{F}$ in \eqref{moy} and the internal energy $U$ in \eqref{uuu} may be written, setting $\beta\equiv 1/\theta$, as:
\begin{align}\label{a}
\ave{F}&=\frac{\partial \ln(Z)}{\beta\, \partial h}\equiv -\frac{\partial A}{ \partial h}\qquad U=-\frac{\partial \ln(Z)}{\partial \beta}\equiv A-\theta \frac{\partial A}{ \partial \theta}\nonumber\\
Z(\beta,h)&=\p \exp(-\beta \,w\,h) -1\q\int_0^\infty dE\, \exp(-\beta \,E)\tau(E/w).
\end{align}
$Z$ is essentially the quantity called in thermodynamics the partition function. It becomes dimensionless if it is divided by the reduced Planck constant $\hbar$, which however plays here no physical role. The continuous energy $E$ in the integral may be replaced by closely-spaced discrete energies $\eps_i,\,i=1,2...$, the spacing between adjacent energies accounting for the $\tau$-function. This procedure is the one employed in the numerical evaluation of integrals; it does not in itself implies quantization. The factor preceding the integral in \eqref{a} entails that the energies $\eps_i,\,i=1,2...$ are multiplied by some function of $h$. In the second expressions of $\ave{F}$ and $U$ given in \eqref{a} we have introduced for brevity the so-called \g{free energy} $A(\theta,h)\equiv -\theta\ln(Z(\theta,\,h))$. 

To be sure, the present paper does not provide explicit expressions of gases internal energy if the law of corpuscle motion remains unknown. It does provide, however, a first-principle proof of the ideal-gas law, including a possible effect of uniform gravity, and the barometric equation. We have recovered the usual thermodynamical and semi-classical statistical-mechanical expressions for the special case of ideal gases.

\section{The energy $\theta$ is the thermodynamic temperature}\label{thermo}

We now prove that $\theta$, introduced in the previous sections on dimensional grounds only, is the thermodynamic temperature. We do this by showing that the maximum efficiency of a thermal cycle employing ideal gases is: $1-\theta_l/\theta_h$, where $\theta_l$ is the cold-bath temperature and $\theta_h$ the hot bath temperature.

From \eqref{a} we obtain:
\begin{align}\label{b}
\ave{F}&= -\frac{\partial A}{ \partial h}\qquad U= A-\theta \frac{\partial A}{ \partial \theta}\nonumber\\
-\delta Q&\equiv dU+\ave{F} \,dh=dA-\frac{\partial A}{ \partial \theta}d\theta-\frac{\partial A}{ \partial h}dh+\theta \,dS=\theta \,dS\qquad S\equiv-\frac{\partial A}{ \partial \theta},
\end{align}
where $\delta Q$ represents the heat released by the gas, from the law of conservation of energy. For any function $f(\theta,h)$ such as $U,\,A,\,S$: $df\equiv \frac{\partial f}{ \partial \theta}d\theta+\frac{\partial f}{ \partial h}dh$. If the gas is in contact with a thermal bath ($\theta$= constant), $\delta Q$ is the heat gained by the bath. The quantity $S$ defined above is called the \g{entropy}. In particular, if heat cannot go through the gas container wall (adiabatic transformation) we have $\delta Q=0$ that is, according to the above result: $dS=0$. Thus adiabatic transformations are isentropic. Note that introduction of $A$ and $S$ is only a matter of mathematical convenience.

A Carnot cycle consists of two isothermal transformations at temperatures $\theta_l$ and $\theta_h$, and two intermediate adiabatic transformations ($dS=0$). After a complete cycle, the entropy recovers its original value and therefore $dS_l+dS_h=0$. According to \eqref{b}: $-\delta Q_l=\theta_l\, dS_l$, $-\delta Q_h=\theta_h\, dS_h$ and therefore $\delta Q_l/\theta_l+\delta Q_h/\theta_h=0$. Energy conservation gives the work performed over a cycle from: $W+\delta Q_l+\delta Q_h=0$. The cycle efficiency is defined as the ratio of $W$ and the heating $-\delta Q_h$ supplied by the hot bath. We have therefore: $\eta\equiv \frac{W}{-\delta Q_h}=\frac{\delta Q_h+\delta Q_l}{\delta Q_h}= 1-\frac{\theta_l}{\theta_h}$, from which we conclude that $\theta$ is the \g{thermodynamic temperature}.

This temperature is defined only to within a multiplicative factor, which is fixed by agreeing that the water triple-point temperature is 273.16 kelvins. One thus sets: $\theta\equiv \kB T$. One generally considers an amount of gas called a \g{mole} occupying a volume of 0.0224 cubic meters at standard pressure and temperature (approximately one atmosphere or 100 000 pascals, and 300 kelvins). We then write: $\textsf{P}\,\textsf{V}=R\,T$, with the ideal-gas constant: $R\approx 8.314...$ joules per kelvin.

\section{Conclusion}\label{conclusion}

Let us briefly recall the concepts introduced in the present paper. One can imagine that after having introduced the corpuscular concept, Democritus observed the elastic bounces of a unit weight on a balance and defined the weight \g{impulse} from the motion period. Not knowing the nature of the motion (parabolic? hyperbolic?), he may have thought of introducing a weight factor such that the average force $\ave{F}$ \emph{does not depend} on the law of motion. This, as we have seen, may be done. This weight factor involves for dimensional reasons a quantity $\theta$ having the dimension of energy. Considering a thermal engine operating between two baths at temperatures $\theta_l,\,\theta_h$ one finds on the basis of the principles just stated that the maximum efficiency is: $1-\theta_l/\theta_h$. This allows us to call $\theta$ the thermodynamic temperature. 

William of Ockham (1287-1347) set as a matter of principle that one should not employ more concepts than those that are strictly necessary to explain the observed phenomenas. (Some authors consider that the Ockham philosophy predates the advent of modern science by insisting on facts and the kind of reasoning employed rather than on speculations about essences). Accordingly, it seems important to elucidate the assumptions on which rest, in particular, the barometric and ideal-gas laws that play an essential role in theoretical physics and many applications. Our thesis is that these laws may be obtain on the sole basis of the Democritus model of corpuscles and vacuum. It is indeed unnecessary to specify the laws of  motion. One can also show that the ideal-gas internal energy depends only on temperature (in the absence of gravity). To evaluate explicitly this energy it is, however, necessary to know the law of motion. From a pedagogical standpoint and in application of Ockham's concept one should not postulate principles which, without being erroneous, are unnecessary.

\appendix

\section{A simple but incomplete proof of the ideal gas law}\label{simple}

We present in this paragraph a simple proof leading to the ideal-gas law. Initially, we only suppose that the gas corpuscles are independent, so that the force exerted on the piston by $N$ corpuscles is $N$ times the force exerted by a single corpuscle, the other conditions (temperature and cylinder height) being the same. We also postulate that the force exerted by a corpuscle of any kind (e.g., nitrogen or helium) on the piston depends only on temperature and height (or volume). At the end of the argument, we additionally postulate that intermediate pistons may be removed without affecting the system equilibrium. This latter assumption would be untenable if the corpuscles were attracting each others, as is the case for non-ideal gases. The present proof is incomplete because we postulate, rather than demonstrate, that plates may be added or removed at various altitudes without resulting into any physical effects.

Let us consider a cylinder of unit area, with a tight piston that can move freely along the axis. This cylinder of height $h$ contains $N$ corpuscles and is raised at some temperature $\theta$.  The cylinder may, for exemple, be filled with nitrogen at standard temperature and pressure. Because the corpuscles are independent, $N$ corpuscles exert $N$ times as much pressure as a single corpuscle, $\theta$ and $h$ being unchanged. We can therefore set in general: $F=N\,f(\theta,h)$, where $f(.,.)$ is an unknown two-variables function. We can (at least in principle) define $\theta$ as the force that a single helium atom exerts on the piston when the cylinder height $h=1$. Then: $F=f(\theta,1)$.

If we now superpose $h$ such cylinders, possibly containing corpuscles of various kinds, the total height becomes $h$ and the number of corpuscles becomes $h$ also. The force $F$ exerted on the upper piston gets transmitted unchanged to each of the cylinders, if we neglect the gas and cylinder weight. This amounts to saying that each cylinder remains in the same conditions as before. But the system presently considered has height $h$, contains $h$ corpuscles, and the force is: $F=\theta$. Substituting these values of $h$, $N$ and $F$ in the general expression: $F=N\,f(\theta,h)$, we get: $\theta=h\,f(\theta,h)$. Thus $f(\theta,h)=\theta/h$, and the general formula becomes: $F=N\,\theta/h$, which is the ideal gas law.

We have implicitly assumes above that one can remove the intermediate pistons without modifying the system equilibrium; this is plausible if the corpuscles do not interact. A proof is given in Appendix \ref{annexe}. \\

\section{Barometric equation with a plate}\label{annexe}

We consider the case where the corpuscle of weight $w$ is bouncing between the ground $z=0$ and a fixed plate at $z=h$. We are seeking the density: $\rho(z_o)$, where $0<z_o<h$. In Section \ref{pot}, we have supposed that this density is the same as if the plate were not there (except for a normalization factor). Even though this assumption is plausible, it is useful to verify it to prove our formalism consistency. As in the main text we denote by $z_m$ the maximum height that the corpuscle would attain if the plate were not there, as a consequence of its energy. We set $w=1,\theta=1$ for brevity. 

To evaluate the probability $p$ that the corpuscle be \emph{below} $z_o$, one must distinguish three cases:
\begin{align}\label{pondbis}
0<z_m<z_o:& \qquad~p=1\nonumber\\
z_o<z_m<h:& \qquad~p=\frac{\tau(z_m)-\tau(z_m-z_o)}{\tau(z_m)}\nonumber\\
z_m>h:&\qquad ~p=\frac{\tau(z_m)-\tau(z_m-z_o)}{\tau(z_m)-\tau(z_m-h)}
\end{align}

On the other hand, the weighting factors relating to $z_m$ given in \eqref{pond} are:
\begin{align}\label{pondter}
z_m<h:&\qquad\exp(-z_m)\tau(z_m)\qquad \nonumber\\
z_m>h:&\qquad\exp(-z_m)(\tau(z_m)-\tau(z_m-h)) 
\end{align}

The probability that the corpuscle be below $z_o$ is therefore:
\begin{align}\label{rhoh}
&\int_0^{z_o}dz_m\,\exp(-z_m)\tau (z_m)+\int_{z_o}^h dz_m\,\exp(-z_m)(\tau (z_m)-\tau (z_m-z_o))\\
&+\int_h^\infty dz_m\,\exp(-z_m)(\tau (z_m)-\tau (z_m-z_o))=(1-\exp(-z_o))\int_0^\infty dz_m\,\exp(-z_m)\tau (z_m).
\end{align}

After normalization, we see that the probability that the corpuscle be below $z_o$ is: $1-\exp(-z_o)$, which is the expected result if the density is the same as in the absence of the plate at $z=h$, namely: $\exp(-z_o)$. 

\section{Non-uniform gravity}\label{gen}

In the present appendix we show how non-uniform weights $w(z)\equiv m\,g(z)$, or continuous potentials: $\phi(z)\equiv\int_0^z dz\,w(z)$, could be handled. On earth, weights decay in proportion to the reciprocal of the square of the distance from the earth center. It is only for small changes in altitude that weights may be considered constant. Another example is a cylinder rotating about its axis at some constant angular rate. In that case $g$ is proportional to the distance from axis.

We only treat the case of two weight values: $w_1,\,0<z≤h$ and $w_2,\,z>h$, and evaluate the ratio of the average times spent by the corpuscle below and above the $h$ altitude. Besides energy conservation we only suppose that corpuscle speeds are continuous. As in the main text, the corpuscles considered are bouncing off the ground at $z=0$, at some temperature $\theta$.

Let us denote by $\tau_i(Z)$ the time-interval corresponding to a distance $Z$ from the top of the trajectory when the weight is a constant $w_i$, with $\tau_i(0)=0$. These functions are of course different for different weight values. Considering two weight-values corresponding to subscripts 1,2, continuity of the corpuscle speed entails that: $\tau_2(Z)=(w_1/w_2)\tau_1(w_2\,Z/w_1)$, as one can see by taking the derivative of this relation with respect to $Z$. This relation is readily verified for the special case of Galilean motion in which case: $\tau_1(Z)=C \sqrt{Z/w_1},~\tau_2(Z)=C \sqrt{Z/w_2}$ for a constant $C$. As in the main text we will let $E$ denote the corpuscle energy.

When $E≤w_1\,h$ the corpuscle stays below the $h$-plane. When $E>w_1h$, the time spent by the corpuscle above the $h$-plane is: $\tau_2(\frac{E-w_1\,h}{w_2})$ because the corpuscle weight is $w_2$ and its energy with respect of the $h$-plane is $E-w_1\,h$. The time spent below the $h$-plane is in that case: $\tau_1(\frac{E}{w_1})-\tau_1(\frac{E}{w_1}-h)$. Thus the average times spent below and above the $h$-plane are respectively:
\begin{align}\label{aaaa}
\textrm{average time below}&=\int_0^{w_1h}\,dE\, \om(E)\,\tau_1(\frac{E}{w_1}) + \int_{w_1h}^\infty\,dE\, \om(E) \p\tau_1(\frac{E}{w_1})-\tau_1(\frac{E}{w_1}-h)\q
    \nonumber\\
    &=\int_0^{\infty}\,dE\, \om(E) \,\tau_1(\frac{E}{w_1})- \int_{w_1h}^\infty\,dE\, \om(E) \,\tau_1(\frac{E}{w_1}-h) \nonumber\\
\textrm{average time above}&= \int_{w_1h}^\infty\,dE\, \om(E) \,\tau_2(\frac{E-w_1\,h}{w_2})=\int_{w_1h}^\infty\,dE\, \om(E) \,\frac{w_1}{w_2}\tau_1(\frac{E}{w_1}-h) .
\end{align}

If $\om(E)$ denotes the energy distribution, the ratio of the average times spent by the corpuscle below and above $h$ is therefore:
\begin{align}\label{h'}
\mathcal{T}=\frac{\int_0^\infty dE\, \om(E) \tau_1(E/w_1)-\int_{w_1\,h}^\infty dE \,\om(E) \tau_1(E/w_1-h)}{(w_1/w_2)\int_{w_1\,h}^\infty dE \,\om(E) \tau_1(E/w_1-h)}
\end{align} 

The only way to remove the $\tau$-functions is to choose: $\om(E)=\exp(-E/\theta)$, where $\theta$ has the dimension of an energy. Proceeding as in the main text, we obtain from the above expression the result:
\begin{align}\label{hh}
\mathcal{T}=\frac{w_2}{w_1}\p\exp(\frac{w_1h}{\theta})-1\q.
\end{align}
We have therefore obtained a result applicable to non-linear potentials. A possible generalization would consist of considering arbitrary static potentials. This, however, will not be done here.

The expression in \eqref{hh} is usually obtained by postulating a gas density: $\rho(z)\propto\exp\p-\phi(z)/\theta\q$, where $\phi(z)$ is the potential at $z$. $\mathcal{T}$ is now viewed as  the ratio of the integral of $\rho(z)$ from $0$ to $h$ and its integral from $h$ to $\infty$. We have: $\phi(z)=w_1\,z,~\,z≤h~  \textrm{and}~\phi(z)=w_1\,h+w_2(z-h)=w_2\,z+(w_1-w_2)\,h,\,z>h$. 

Thus: $ \rho(z)=C \exp\p -\frac{w_1\,z}{\theta}  \q,\,z≤h, ~\textrm{and}~\rho(z)=C \exp\p -\frac{w_2\,z+(w_1-w_2)\,h}{\theta}  \q,\,z>h$, where $C$ is a constant. Integration gives
\begin{align}\label{hhh}
\mathcal{T}_{\textrm{Boltzmann}}&=\frac{\int_0^h dz\,\exp(-w_1\,z/\theta)}{\exp\p (w_2-w_1)h/\theta\q\int_h^\infty dz \exp(-w_2\,z/\theta)}\nonumber\\
&=\frac{w_2}{w_1}\p  \exp(\frac{w_1h}{\theta})-1    \q
\end{align}
which coincides  with \eqref{hh}. The interest of our method is of course that the Boltzmann factor needs not be postulated.

\section{General equations of motion}\label{motion}

The Hamilton equations of motion in one space dimension ($z$) derive from a hamiltonian function: $H(p,z)$, with $v(t)\equiv \frac{dz(t)}{dt}= \frac{\partial H(p,z)}{\partial p},~ \frac{dp(t)}{dt}= -\frac{\partial H(p,z)}{\partial z}$. In the situations considered in this paper we may set $H(p,z)=F(p)+\phi(z)$, where $F(.)$ is a (nearly) arbitrary function of $p$, and $\phi(.)$ is the potential function. In a static (time-independent) force of constant magnitude $w$ directed along the negative $z$-axis, we have $\phi(z)=w\,z$. Thus, a general equation of motion is: $dz=f(p)\,dt$, where $f(p)\equiv \frac{dF(p)}{dp}$, and $p=-w\,(t-t_o)$, where $t_o$ is an integration constant. 

In particular, for a slow corpuscle of mass $m$ we have: $f(p)=p/m$. Since $w=m\,g$, where $g$ is the acceleration, it follows that $z=-\frac{1}{2}g\,t^2$, if $t_o=0$, according to Galileo. For a corpuscle at any speed we have: $f(p)=\frac{p/m}{\sqrt{1+(p/mc)^2}}$ where $c$ is the speed of light, according to Einstein. Still other forms, which are dimensionally correct but unlikely to be in general physically significant, would be: $f(p)=c\,g(p/mc)$ for some function $g(.)$. No particular $g(.)$ function is presumed in this paper.


\end{document}